\documentclass[sigconf,natbib=false]{acmart}

\AtBeginDocument{%
  }

\setcopyright{acmcopyright}
\copyrightyear{2024}
\acmYear{2024}
\acmDOI{XXXXXXX.XXXXXXX}

\RequirePackage[
  datamodel=acmdatamodel,
  style=acmnumeric,
  ]{biblatex}

\usepackage{tikz,subcaption}
\usetikzlibrary{positioning}
\usepackage{color}
\usepackage{soul}
\begin{document}

\title{Weighted Regression with Sybil Networks}


\author{Nihar Shah}
\email{nihar@mystenlabs.com}
\affiliation{%
  \institution{Mysten Labs}
  \country{}
}

\renewcommand{\shortauthors}{Shah}

\begin{abstract}
In many online domains, Sybil networks -- or cases where a single user assumes multiple identities -- is a pervasive feature. This complicates experiments, as off-the-shelf regression estimators at least assume known network topologies (if not fully independent observations) when Sybil network topologies in practice are often unknown. The literature has exclusively focused on techniques to \textit{detect} Sybil networks, leading many experimenters to subsequently exclude suspected networks entirely before estimating treatment effects. I present a more efficient solution in the presence of these suspected Sybil networks: a weighted regression framework that applies weights based on the probabilities that sets of observations are controlled by single actors. I show in the paper that the MSE-minimizing solution is to set the weight matrix equal to the inverse of the expected network topology. I demonstrate the methodology on simulated data, and then I apply the technique to a competition with suspected Sybil networks run on the Sui blockchain and show reductions in the standard error of the estimate by 6 - 24\%.
 \end{abstract}



\settopmatter{printfolios=true}
\maketitle

\section{Introduction}
\label{sec:introduction}
The Sybil problem -- where an individual has multiple accounts -- is a well-known problem on many online platforms, since first being formalized in \cite{douceur2002sybil}. Depending on the setting, this can be forbidden, tolerated, or encouraged; and this can be a minor or pervasive problem. For instance, Facebook (operated by Meta) strictly insists in its terms of service that each user have only one account, while Instagram (also operated by Meta) explicitly allows users to have multiple identities. These problems are most extreme in crypto, where an account is entirely derived from a seed phrase of several ordered words; and so a single user creating thousands of accounts is feasible and potentially worthwhile, especially in anticipation of monetary rewards.

The Sybil problem complicates the analysis of experiments. Most estimators assume independent observations; or at best adjust for possible dependence in the variance-covariance matrix via clustering \cite{qjac038}. These are both insufficient for the Sybil problem. In practice, therefore, experimenters often attempt to detect Sybil networks prior to the experiment and exclude suspected cases outright (or collapse them into a single observation). This is a reasonable approach when the experimenter has very high confidence that a set of observations is linked. However, in cases where that confidence is lacking, such an approach can be inefficient and dilutes the power of the experiment.

This paper addresses precisely this scenario, where the probability that a set of observations is a Sybil network is neither zero nor one; and so neither full exclusion nor full inclusion is the optimal solution. I frame the question as a weighted regression problem, and show that the optimal weight matrix -- to minimize the mean-squared error of the estimator -- is simply the inverse of the expectation of the network topology across different permutations. This solution nicely handles many practical cases, such as the possibility of several Sybil networks either being distinct networks or being one large network. For the most common setup -- multiple distinct potential Sybil networks whose observations have homoskedastic errors -- I also offer a direct solution that does not require matrix inversion, so that this methodology can be deployed successfully on large populations without hitting computational limits.

This is a natural extension from the existing literature, which focuses heavily on identifying Sybil networks either through ad-hoc analyses, e.g. \cite{yangwilson}, or through systematic approaches, e.g. \cite{danezis2009sybilinfer} or \cite{6787042}. These different approaches can help experimenters find potential networks and assess their likelihoods of being interlinked or independent observations. At that point, the methodology in this paper can be applied to use that information efficiently, downweighting suspected observations by the probability and potential size of the network.

After demonstrating with some simulated data (where the data-generating process is of course known), I show an application in a real-world setting. In late fall 2023, Mysten Labs held a competition on the Sui blockchain, where participants could play on-chain games to win prizes. Given the monetary stakes, we suspect some of those players created multiple identities. This confounds our ability to estimate the long-run responsiveness of participants' usage to unanticipated bonus rewards. However, we are able to identify potential Sybil networks; and I can therefore apply this paper's methodology to the data accordingly. The results show a drop in the standard error of the estimated treatment effect of 6\% and 24\%, relative to the baseline approaches of full inclusion or full exclusion respectively.

The paper proceeds as follows. Section \ref{sec:model} establishes the optimal weights for this problem, both in the general case and in a simplified case with computational limits. Section \ref{sec:simulation} demonstrates an application with simulated data, where the data-generating process is known. Section \ref{sec:empirical} considers the application on real data, regarding the competition on the Sui blockchain with suspected Sybil actors. Section \ref{sec:conclusion} concludes.

\section{Optimal Weights}
\label{sec:model}
This section establishes the optimal weights for the MSE-minimizing weighted regression estimator. This optimal weight matrix is the inverse of the expected graph topology. I also simplify this further for the most common case, where there are multiple disjoint possible Sybil networks, such that no matrix inversion is necessary.

First, I use a simple model to build intuition. Consider a setting with $n_1 + n_2$ observations, where the first $n_1$ observations are independent and the remaining $n_2$ observations (in set S) are either independent with probability $1 - \pi$ and in a Sybil network with probability $\pi$. For further simplicity, assume homoskedastic errors with variance $\sigma^2$, a single explanatory variable whose mean is zero, and a single response variable whose mean is zero. In this simplified setting, the weighted linear regression estimator for $\beta$ is the summation of weighted $xy$ divided by the summation of weighted $x^2$. I thus want to find the weights to minimize the mean-squared error of the estimator.
$$
w_i^* = \text{arg}\min_{w_i} \mathbb{E}\left(\hat{\beta} - \beta\right)^2 = \mathbb{E}\left(\frac{\sum_i w_i x_i y_i }{\sum_i w_i x_i^2} - \beta\right)^2
$$

This can be simplified, e.g. by removing the denominator of the first term and instead imposing a condition on the sum of weights, to the following equation. The cross-products stem from the dependence between the errors in $S$ if they form a Sybil network.
$$
w_i^* = \text{arg}\min_{w_i}  \sum_i w_i^2 x_i^2 \sigma^2 + \pi \sum_{i \in S} \sum_{j \in S, i \neq j} w_i w_j x_i x_j \sigma^2,~\sum w_i = 1
$$

Finally, $w_i$ is set to $w_1$ (corresponding to the independent observations) and $w_2$ (corresponding to the possible Sybil observations). This yields an optimal solution in Equation \eqref{eq:simple} for the ratio of weights. Under the admittedly tenuous assumption that the sample averages of the squares and cross-products of $x$ are approximately equal, this simplifies further.
\begin{equation}
\frac{w_1^*}{w_2^*} = \frac{\mathbb{E}(x^2|S) + \pi (n_2 - 1) \mathbb{E}(x_ix_j|S, i \neq j)}{\mathbb{E}(x^2|\bar{S})} \approx 1 + \pi (n_2 - 1)
\label{eq:simple}
\end{equation}

Equation \eqref{eq:simple} is intuitive. When $\pi = 0$, i.e. there is definitively no Sybil network, then the observations are weighted equally. When $\pi = 1$, i.e. there is definitively a Sybil network, then the Sybil observations are effectively rolled into a single one by individually being weighted down by $n_2^{-1}$, relative to other observations. Finally, the observations are weighted down by some intermediate value when $0 < \pi < 1$.

I now turn to the general case. This leverages the insight that any weight estimator in the following class of linear estimators is unbiased for a well-defined weight matrix $W$. As such, minimizing the mean-squared error for $\hat{\beta}$ is the same as minimizing its variance.
$$
\hat{\beta} = \left(x^T W x\right)^{-1} x^T W y
$$

There are two sources of uncertainty for $\hat{\beta}$: the errors $\epsilon$ (assumed to be homoskedastic) and the network topology (where each possible topology is denoted $G_i$). Thus, I apply the law of total variance to the estimator and proceed to simplify it across four steps.
\begin{align*}
\mathbb{V}\left(\hat{\beta}\right) =& \mathbb{E}\left(\mathbb{V}\left(\hat{\beta} | G\right)\right) + \mathbb{V}\left(\mathbb{E}\left(\hat{\beta} | G\right)\right) = \sum_i \pi_i \mathbb{V}\left(\hat{\beta} | G\right) \\
=& \sum_i \pi_i (x^T W x)^{-1} x^T W \Sigma_i W x (x^T W x)^{-1} \\
=& \sum_i \pi_i \sigma^2 (x^T W x)^{-1} x^T W G_i W x (x^T W x)^{-1} \\
=& \sigma^2 (x^T W x)^{-1} x^T W \left(\sum_i \pi_i G_i\right) W x (x^T W x)^{-1}
\end{align*}

This can be explained as follows. In the first step, the variance of the expectation term can be dropped due to the unbiased property of the estimator; and the second step writes out the expectation of the variance term. The third step uses the fact that any given network topology $G_i$ is equivalent (up to the constant $\sigma^2$) to that topology's variance-covariance matrix $\Sigma_i$. The topology matrix $G$ is a symmetric matrix of zeroes (for independent observations) and ones (for identical observations), and this is precisely the correlation matrix of errors too for the given topology. The fourth step simply leverages the distributive property of matrix multiplication.

At this point, I use the well-known result that the weight matrix should be set to the inverse of the inner term to minimize the expression, e.g. \cite{carroll2017transformation}. This yields Equation \eqref{eq:general}, where the optimal weight matrix is the inverse of the expected network topology.
 \begin{equation}
W^* =  \left(\mathbb{E} G \right)^{-1}
\label{eq:general}
\end{equation}

Indeed, Equation \eqref{eq:general} holds in the presence of heteroskedasticity or even arbitrary variance-covariance matrices, as long as either the elements of $G$ are adjusted correctly or instead the inverse of $\mathbb{E}\Sigma$ is directly used instead. Separately, note that Equation \eqref{eq:general} assumes the expected topology is invertible. If there is \textit{guaranteed} Sybil network that exists across all possible network topologies, this matrix is not full rank and thus not invertible. The solution is simply to remove or roll up the guaranteed network; or alternatively to use other possible inversion techniques (e.g. the Moore-Penrose inverse) that are robust to deficient-rank matrices.

Equation \eqref{eq:general} does not scale easily to settings with many observations, as matrix inversion is a computationally expensive endeavor. However, for the most common setting -- a set of distinct potential networks, where each potential Sybil network $s$ is the same actor with $\pi_s$ probability and contains $n_s$ observations, the optimal weight matrix can be constructed directly, bypassing the need for the inversion. Each entry of the matrix has the following form in Equation \eqref{eq:matrix}. This also includes observations that are not in any network and thus independent, where $\pi_s = 0$ and $n_s = 1$.
\begin{equation}
w(i,j)^* = \begin{cases} \frac{1 + (n_s - 2) \pi_s}{(1 - \pi_s)(1 + (n_s - 1)\pi_s)}~\textit{if}~i=j, i \in~\text{network}~s \\
\frac{-\pi_s}{(1 - \pi_s)(1 + (n_s - 1)\pi_s)}~\textit{if}~i\neq j~\text{and}~i, j \in~\text{network}~s \\
0~\text{otherwise}
 \end{cases}
 \label{eq:matrix}
\end{equation}

\section{Simulation}
\label{sec:simulation}
I demonstrate the efficacy of the methodology in simple setting with simulated data. In particular, the data-generating process has the following form.
$$
y_i = 2 + 3 x_{1,i} - \frac{1}{2} x_{2,i} + \epsilon_i~~\text{where}~\epsilon \sim N(0, 1)
$$

In this setting, there are 400 observations and up to seven distinct potential Sybil networks with probabilities between 0.1 and 0.9 of being Sybil and sizes between 10 and 90 observations. I simulate the setting 1000 times, and compare across six methods for generating $\hat{\beta}$, scoring by the mean-squared error between $\hat{\beta}$ and the known $\beta$. The six methods are: an estimator that excludes all suspected Sybil networks (``exclusion''), an estimator that includes all observations (``inclusion''), an estimator that excludes suspected networks with probabilities greater than 50\% of being Sybil (``threshold''), an estimator that repeatedly re-samples entire networks by their probabilities of not being Sybil (``network-sampled"), an estimator that samples individual observations by their probabilities of not being Sybil (``observation-sampled"), and the optimal weighted estimator from Equation \ref{eq:general} (``weighted'').

\begin{figure}
\caption{Mean-Squared Error by Estimator, Simulation}
\includegraphics[scale=0.3]{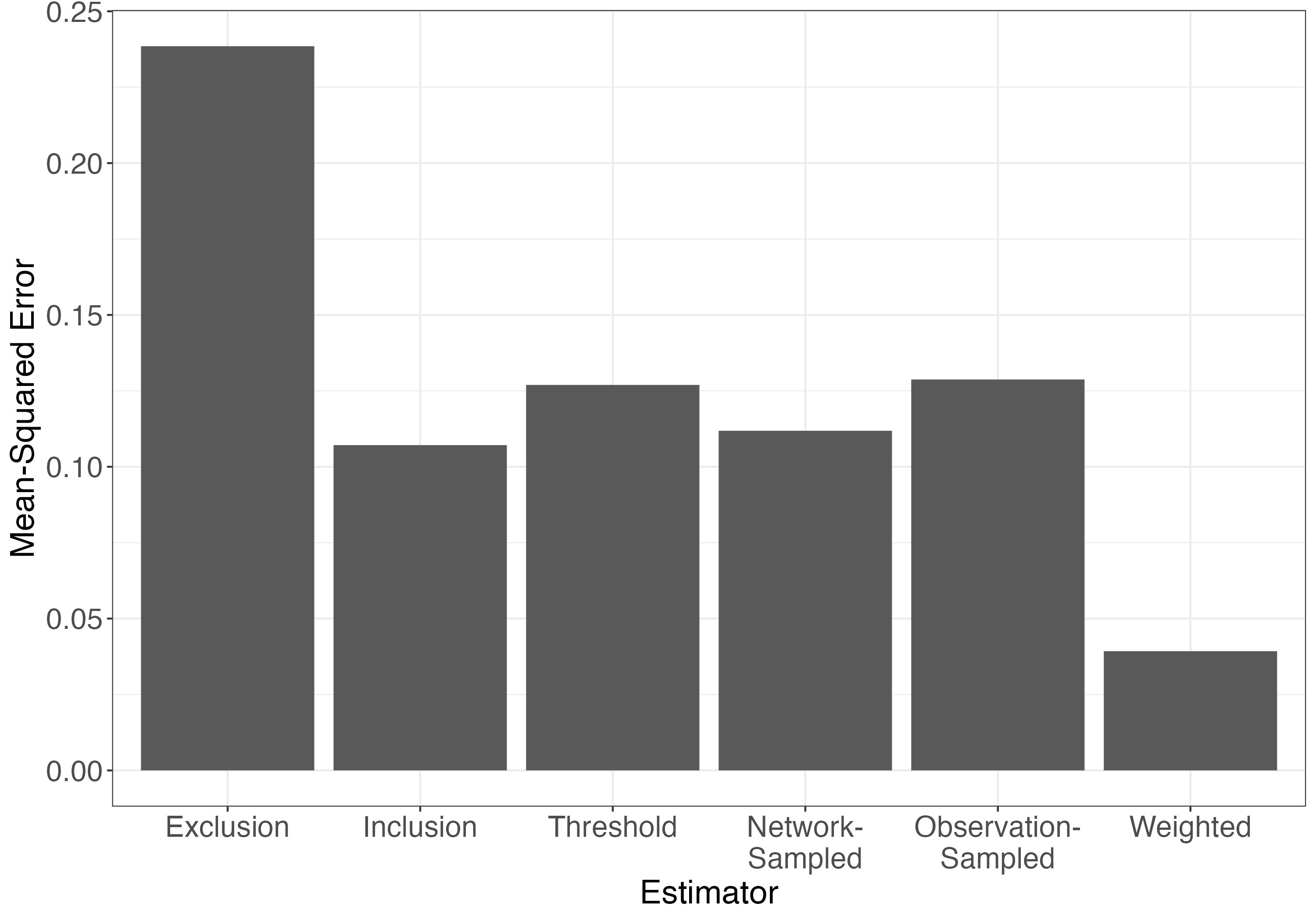}
\label{fig:simulation}
\end{figure}

Figure \ref{fig:simulation} shows the results. The exclusion-based estimator performs poorly, given how many observations it excludes. The other four benchmark estimators perform comparably, choosing different approaches to the trade-off between sample size and error correlation. However, the weighted estimator has markedly better performance in mean-squared error from the true coefficient value, compared to all other estimators. This is potentially especially surprising when compared to the estimators that sample networks and observations. However, the weight matrix is not a simple diagonal matrix that downweights suspected members of Sybil networks by probabilities and sizes; and it has off-diagonal elements that further adjust the estimates. Those are critically important, empirically.

This simple check demonstrates the potential value of the approach. By weighting potentially Sybil networks correctly and neither fully including nor excluding them, the estimator is more efficient.

\section{Empirical Application}
\label{sec:empirical}
In late October and November 2023, Mysten Labs held a competition on the Sui blockchain known as ``Quest 3.'' This serves as a good use case for the methodology, given the experimental design and the existence of Sybil networks in the data.

As background, the Quest 3 competition allowed users to play games associated with the Sui blockchain and win prizes (Sui tokens) on the basis of their scores. At the end of the competition, three adjustments were made to the list of eligible participants and prizes. First, the developers of the games that participated in Quest 3 flagged certain participants for inorganic behaviors and their scores were deducted points, causing the list to be re-ranked. Second, wallets demonstrating obvious forms of inorganic behavior in entering the competition, explained shortly, were deemed ineligible. Third, the prizes were made more slightly generous than communicated, largely by extending the cutoffs in ranks for each tier of rewards. This combination, detailed in \cite{mysten}, jointly delivered a set of orthogonal shocks to the system, causing participants to receive potentially different prizes than the ones they were anticipating and setting the stage for an experimental analysis. Specifically, in the thirteen weeks following Quest 3, the prize-eligible participants were measured on the number of days (out of a total of 91) that they took an action on the blockchain. This measure of retention was regressed against the incremental rewards, and the associated parameter estimate could be interpreted as a customer acquisition cost: for every extra unexpected Sui token provided, how many extra days did a participant return to the Sui blockchain?

Crypto, however, remains particularly vulnerable to the Sybil attack, given the ease of creating new addresses and the high-powered incentives to do so. For instance, \cite{artemis} estimates that on certain applications (e.g. the zkSync bridge), almost 50\% of the wallets using the application could represent Sybil activity. Quest 3 also had Sybil activity. In particular, the most Sybil-prone component to Quest 3 was its referral program, as one way to join Quest 3 was to be referred by a participant in the program. However, wallets using this pathway also created the trail needed to identify the networks, through these referral patterns.

\begin{figure}
\centering
\caption{Sample Referral Path}
\begin{tikzpicture}[
    node distance = 0.5cm,
    every node/.style = {circle, draw, minimum size=0.2cm}
]
\node (A) {$W_2$};
\node[left=of A] (B) {$W_1$};
\node[right=of A] (C) {$W_3$};
\node[below right=of C] (D) {$W_4$};
\node[above right=of D] (E) {$W_5$};
\node[below right=of E] (F) {$W_6$};
\node[right=of E] (G) {$W_7$};
\node[right=of F] (H) {$W_8$};
\draw[->] (B) -- (A);
\draw[->] (A) -- (C);
\draw[->] (C) -- (D);
\draw[->] (C) -- (E);
\draw[->] (E) -- (F);
\draw[->] (E) -- (G);
\draw[->] (F) -- (H);
\end{tikzpicture}
\label{fig:referral}
\end{figure}
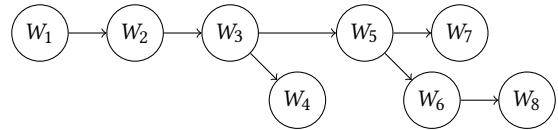

Figure \ref{fig:referral} demonstrates one such example. A wallet would refer a wallet or set of wallets, which would in turn refer another wallet or set of wallets. Organic trees would be relatively short and thin, but Sybil trees would be wide and deep. For example, one individual created a network of over 156,000 wallets that referred one another. This and other egregious cases were indeed removed from the competition and not given rewards, as explained in \cite{mysten}. However, there were several marginal cases that were ultimately included and potentially given rewards. Their possible presence as Sybil networks confounds our ability to estimate treatment effects.

I thus identify potential Sybil networks in two ways. First, I find referral trees that had twenty or more individuals. While some of these could represent organic behavior by social users, others would surely represent inorganic behavior by Sybil users. Second, I find cases where users in a referral tree had transferred money to one another ten or more times. (In order to play games in Quest 3, users need a small stock of tokens -- often only a few pennies -- to pay transaction costs. Independent users have independent sources of funds, but Sybil networks must seed their wallets.) Finally, these potential networks are assumed to have a probability of being Sybil as 50\%. This is a crude approximation, but without ground truth on Sybil activity, this is the best one can do.

I estimate treatment effects under three models: the inclusion estimator, the exclusion estimator, and the weighted estimator. The standard errors for the weighted estimator can be computed easily, in Equation \eqref{eq:variance}, because of the equivalence (up to a constant) of the expected network topology and the expected variance-covariance matrix.
\begin{equation}
\label{eq:variance}
\mathbb{V}\left(\hat{\beta}\right) = \hat{\sigma}^2 \left(X^T W X\right)^{-1}
\end{equation}

\begin{figure}
\caption{Standard Error by Estimator, Quest 3}
\includegraphics[scale=0.3]{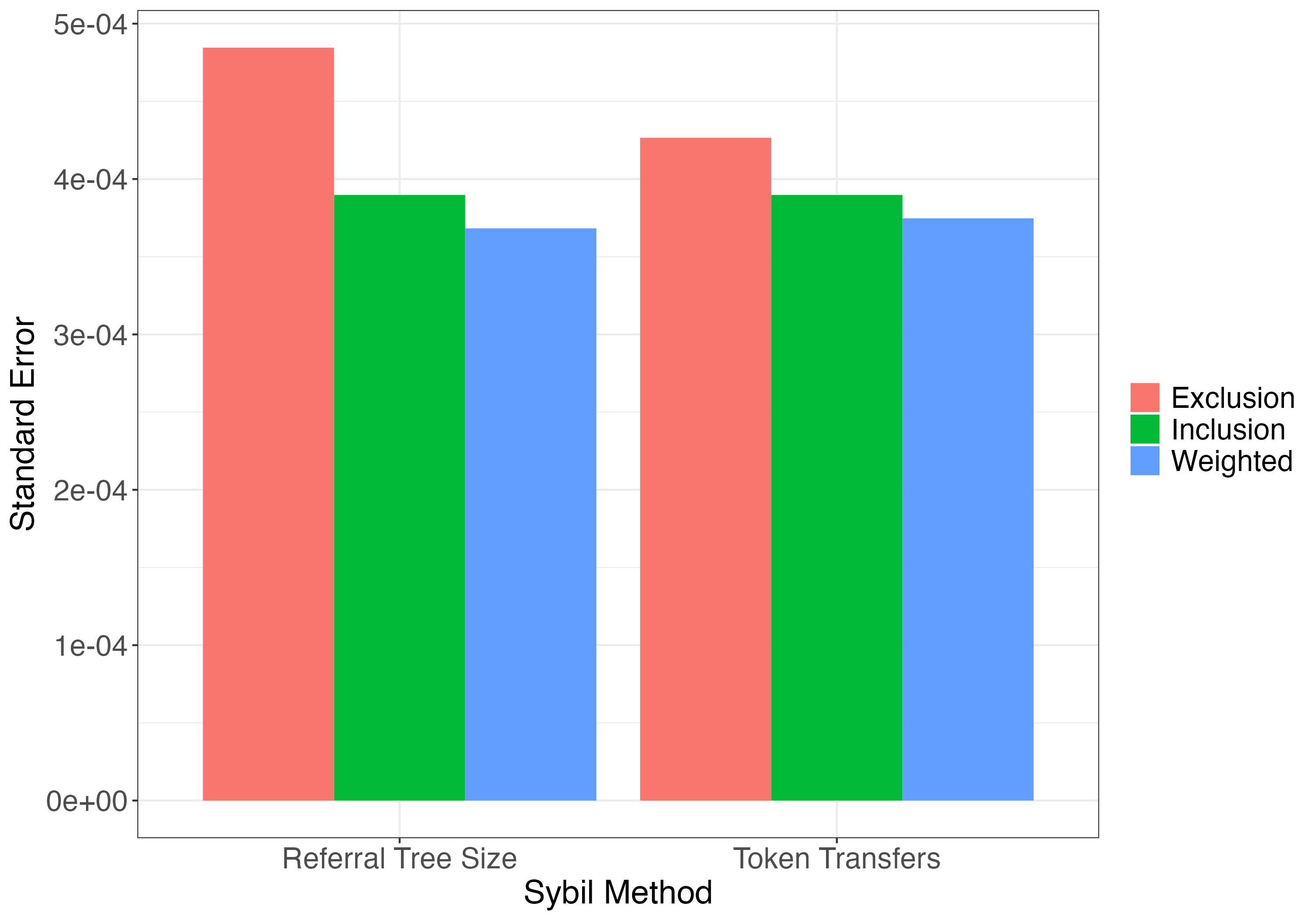}
\label{fig:quest3}
\end{figure}

The results are in Figure \ref{fig:quest3}. The treatment effects are not reported publicly due to business sensitivity, but they are virtually identical under all three models. However, the standard errors are lowest for the weighted estimator relative to the other two. Specifically, the weighted estimator finds a standard error that is 24\% and 6\% smaller than those of the exclusion and inclusion estimators respectively, for Sybil networks identified from the referral tree size. The weighted estimator also delivers gains for Sybil networks identified from token transfers, with standard errors that are 12\% and 4\% smaller than those of the exclusion and inclusion estimators.

\section{Conclusion}
\label{sec:conclusion}
Sybil activity is common in online settings and especially in crypto settings. The current approaches -- which often involve full inclusion or exclusion of suspected Sybil networks, in an ad-hoc way -- are inefficient and dilute the power of an experiment. This paper makes better use of the data by formulating the problem as a weighted regression and constructing optimal weights as the expectation of the different candidate network topologies. This shows promise both in simulated data (where the true treatment effects are known) and in real-world data.

But there is more to do, including better defenses against Sybil vulnerabilities, richer ways of modeling Sybil network probabilities, and -- above all -- more acknowledgement of their presence in online settings. Sybil networks are not going away; and rather than ignore them, experimenters should explicitly incorporate their presence into models and build robust estimation tools.

% \begin{acks}
% acksacksacks
% \end{acks}

\newpage
\printbibliography

% \appendix

% appendixappendixappendix

\end{document}